\documentclass[12pt]{article}

\usepackage[dvipdfm]{graphicx}
\begin{document}

\title{ Digital herders and    phase transition  \\
in \\
a voting model}

\author{M Hisakado\footnote{[1]
masato\_hisakado@standardandpoors.com}  \space{} and   S Mori\footnote{[2] mori@sci.kitasato-u.ac.jp} }

\maketitle

*Standard  \& Poor's, Marunouchi 1-6-5, Chiyoda-ku, Tokyo 100-0005, Japan

\vspace*{1cm}

\dag Department of Physics, School of Science,
Kitasato University, Kitasato 1-15-1 , Sagamihara, Kanagawa 228-8555, Japan

   \vspace*{1cm}

\begin{abstract}
In this paper, we discuss a voting model with two candidates, $C_1$ and  $C_2$.
We set two types of voters--herders and independents.
The voting of independent voters  is  based on  their fundamental values;
on the other hand, 
the   voting of herders is  based on the number of votes.
Herders always select the majority of the previous $r$ votes, which is visible to them.
 We call them   digital herders. 
We can accurately  calculate the distribution  of votes   for  special cases.
When  $r\geq 3$,  we find that  a  phase transition  occurs at  the upper limit of  $t$,  where $t$ is the discrete time (or  number of  votes). 
As the fraction of herders increases, the model features   a phase transition beyond which  a state where most voters make the correct  choice coexists with one where most of them are wrong.
On the other hand, when  $r<3$, there is no phase transition.
In this case, the herders' performance is the same as that of the independent  voters.  
Finally,  we recognize  the behavior of  human beings by conducting  simple experiments.

\end{abstract}



\newpage
\section{Introduction}

In general collective herding  poses  interesting problems in several fields. To cite  a few examples in statistical  physics, anomalous fluctuations in   financial markets \cite{Cont}\cite{Egu} and  opinion dynamics \cite{Stau}  have been related  to percolation and the  random field Ising model. 
To estimate public perception,  people observe the actions of other individuals; then, they  make a  choice similar to  that of  others. 
Recently,  these behaviors have been referred to  as Kuki-wo-yomu (follow an atmosphere) in Japanese.    
 Because it is usually sensible to do what other people are doing, the phenomenon is assumed to be the result of  a rational choice. Nevertheless, this approach can sometimes  lead to arbitrary or even erroneous decisions. This phenomenon  is known as an  information cascade \cite{Bikhchandani}.

A recent agent-based model proposed  by Curty and Marsili \cite{Curty} focused on the limitations  imposed by herding on the efficiency of  information aggregation. Specifically, it was shown that when the fraction of herders in a population of agents increases, the probability that herding yields  the correct forecast (i.e.,  individual information bits are correctly aggregated) undergoes a transition to  a  state in which either all herders  forecast  rightly or no herder does.  

In the previous paper, we introduced a voting model that is similar to  a Keynesian beauty contest  \cite{Keynes}\cite{Hisakado2}\cite{Mori2}.
There are two types of voters--herders and independents--and two candidates.
Herders are known  as copycat voters; 
they vote for  each candidate with the probabilities that are proportional  to the candidates' votes.
In the previous paper, they were known as analog herders.
We investigated a case wherein  all the  voters are herders \cite{Mori}.
In such  a case, the process is a P\'{o}lya process, and the voting rate converges to a beta distribution in a large time limit \cite{Hisakado}.
Next, we doped independent voters in herders.
The proposed 
 voting model is a binomial distribution  doped in a beta binomial distribution mathematically.
In the upper limit of $t$, the  independent voters make the distribution of votes converge to Dirac measure against herders. 
This model  consists of   three phases.
If herders constitute  the   majority or even  half of the total voters,
the voting rate  converges  more  slowly   than  it would  in a  binomial distribution.
If independents constitute the majority of the  voters,
the voting rate converges   at the same rate  as it would  in a  binomial distribution. 
The  phases differ  in terms of  the velocity of the convergence.
If the independent voters vote for the correct candidate rather  than for the wrong  candidate,  the model consists of  no case wherein the majority of the voters select  the  wrong answer. 
The herders affect only   the speed of the convergence;  they   do not affect the  voting rates  for  the  correct candidate. 

The model introduced by Curty and Marsili has  a  limitation  similar to that  of our previous  model in the case wherein  voters  are unable to see  the votes of  all the voters; they  can only  see   the votes of previous voters. However, there is a  significant  difference between our model and their model with  respect to the  behavior of the  herders.  In their model, the  herders always  select the majority of  the votes, which  is visible to them.  Thus,  their  behavior becomes      digital (discontinuous). 
Digital herders have  a stronger  herding power than  analog herders.

Here,  we  discuss  a  voting model  
with two candidates, $C_0$ and $C_1$.
We set two types of voters--independent and herders.
In this paper, the  herders are    digital herders, as in the case of  the model introduced by Curty and Marsili.
The voting of independent voters  is  based on  their fundamental values.
On the other hand, 
the voting of herders  is  based on the number of votes.
Herders always select the majority of the previous $r$ votes, which is visible to them.

The remainder  of this paper is organized  as follows.
In section 2, we introduce our  voting model, and 
 we mathematically 
 define the two types of  voters--independents and herders. 
The voters can see the  previous $r$ votes of  the voters. 
In section 3,
we  calculate the  exact distribution functions of the votes    for the case
 wherein  the voters can see the votes of all the voters.
We discuss the phase transition using the exact solutions.
In section 4, we discuss 
 the special case, $r=1$.
In this case, we calculate the exact   distribution function; however,
there is no phase transition.
In section 5, we analyze the model using mean field approximation.
We  can show   that the phase transition in this system occurs when $r\geq 3$. 
In  section 6,  we describe numerical simulations performed  to confirm the  analytical results pertaining to the asymptotic behavior.
In section 7, we conduct  simple social experiments to recognize the   behavior of human beings.
Finally, the conclusions are presented in section 8.

\section{Model}

We model the voting of two candidates, $C_0$ and $C_1$;
at  time $t$, they have  $c_0(t)$ and $c_1(t)$ votes, respectively.
At each time step, one  voter  votes   for one  candidate;
 the voting is sequential.
Voters are allowed to see  $r$ previous votes for each candidate when they vote  so that they are aware of   public perception.
If $r>t$, voters can see  $t$ previous votes for each candidate.
 At  time $t$, the number of votes for $C_0$ and $C_1$ are $c_0^{r}(t)$ and $c_1^{r}(t)$, respectively.
In the limit $r\rightarrow \infty$,  voters can see  all previous votes. Therefore, $c_0^{\infty}(t)=c_0(t)$ and $c_1^{\infty}(t)=c_1(t)$.

There are  two types of voters--independents and herders;
 we assume   an  infinite number of voters.
Independent voters vote for $C_0$ and $C_1$
with  probabilities $1-q$ and $q$, respectively.
Their votes are independent  of  others' votes, i.e.,
their votes are based  on  their fundamental values.
Here, we set $C_0$ as the wrong  candidate and $C_1$ as the correct candidate to validate the performance of the  herders.
We can  set $q\geq 0.5$  because we  believe that    independent voters vote for the correct  candidate $C_1$ rather   than for the wrong candidate $C_0$.
In other words, we assume that  the  intelligence of the independent voters is virtually   correct.
   
\begin{figure}[h]
\includegraphics[width=120mm]{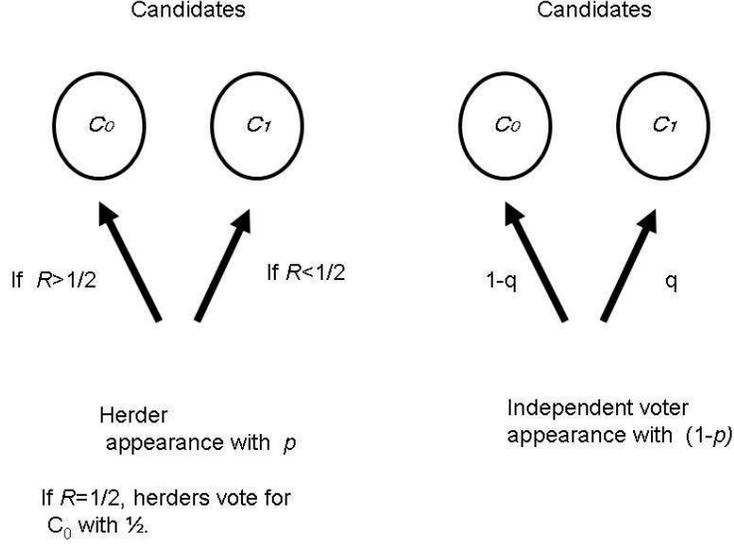}
\caption{Demonstration of  model. $R=c_{0}^{r}/\{c_{0}^{r}+c_{1}^{r}\}$.}
\label{model}
\end{figure}

On the other hand, herders vote for a  majority
candidate;
if $c_0^{r}(t)>c_1^{r}(t)$, herders vote for the candidate $C_0$.
If $c_0^{r}(t)<c_1^{r}(t)$, herders vote for the candidate $C_1$.
If  $c_0^{r}(t)=c_1^{r}(t)$, herders vote for $C_0$ and $C_1$
with the same  probability, i.e.,$1/2$.
In the previous paper, the  herders   voted for each candidate with  probabilities that were  proportional to the candidates' votes \cite{Hisakado2};
they were known as  analog herders.
On the other hand, the   herders in this paper are known  as digital  herders (Fig. \ref{model}).

The independent voters and herders appear randomly and vote.
We set the ratio of independent voters to 
herders as $(1-p)/p$.
In this paper we mainly pay attention in large $t$ limit.
It means the voting of infinite voters.

\section{Exact solutions for $r = \infty$}

In this section, we study the exact solution of the case  $r=\infty$ by
using combinatorics.

Here, we  map the model to  correlated  Brownian motion along the edges of a grid with square cells,  and  we  count the number of  paths. 
Let $m$ and $n$  be the horizontal axis and  the  vertical axis, respectively.
The coordinates of the lower left corner are  $(0,0)$;
 this is the starting point.
$m$ is the number of  voters who vote for  $C_{1}$, and
$n$ is the number of  voters who vote  for  $C_{0}$.
A path shows the history of the votes. 
If a voter   votes  for $C_{1}$, the path move rightwards.
If a voter    votes  for $C_{0}$, the path move upwards.  

\begin{figure}[h]
\includegraphics[width=120mm]{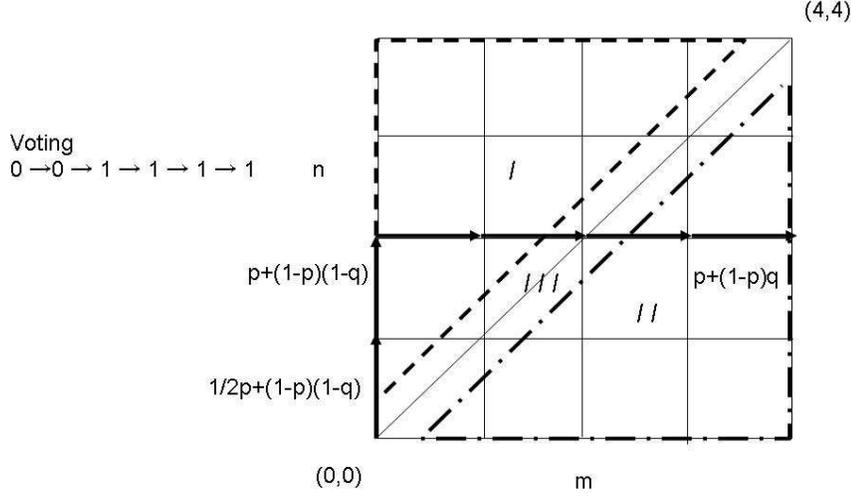}
\caption{Voting and Path. A path shows the history of the votes. 
If a voter   votes  for $C_{1}$, the path move rightwards.
If a voter    votes  for $C_{0}$, the path move upwards.
The sample arrowed line shows the voting of 6 voters, $0,0,1,1,1,1.$ We refer to the regions $m<n$, $m>n$, and $m=n$ as $I$, $II$, and $III$ respectively.  }
\label{f2}
\end{figure}

We define $P_{i}(m,n)$ as the probability    that the   $(n+m+1)$th voter votes
for the candidate $C_i$, where $i=0,1$.
The probability   of moving upwards  is as follows. 
\begin{equation}
P_{0}(m,n)=\left\{ 
\begin{array}{lll}
p+(1-p)(1-q)\equiv A& m<n; \nonumber \\
\frac{1}{2}p+(1-p)(1-q) \equiv B& m=n; \nonumber \\
 (1-p)(1-q) \equiv C& m>n. 
\end{array} 
\right \}
\label{prob}
\end{equation} 
The probability of moving rightwards is $P_{1}(m,n)=1-P_{0}(m,n)$ for each case.
Here, we introduce $X(m,n)$ as the probability that the path passes through  the point $(m,n)$.
The master equation is 
\begin{equation}
X(m,n)=P_{1}(m-1,n)X(m-1,n)+P_{0}(m,n-1)X(m,n-1),
\end{equation} 
for $m\geq 0$ and $n \geq 0$, with the initial condition $X(0,0)=1$.
This defines $X(m,n)$ uniquely.
Hereafter, we refer to the region $m<n$  as $I$, $m>n$ as  $II$, and 
$m=n$ as $III$ (Fig. \ref{f2}).

First, we consider the case $q=1$.
At this limit, independent voters 
always vote for only one candidate, $C_1$
(if we set $q=0$, independent voters vote only for  $C_0$).
The probability is reduced from (\ref{prob}) to 
\begin{equation}
P_{0}(m,n)=\left\{ 
\begin{array}{lll}
p& m<n;\\
\frac{1}{2}p& m=n; \\
0& m>n. \\
\end{array} \right \}
\label{p}
\end{equation} 
In this case,  if the path enters  $II$ ($m>n$), it  can only  move   rightwards. Hence,  $n=m-1$ becomes the absorption wall,  where $m,n\geq0$.
There is a difference between the probability (\ref{p})  in  $I$ ($m<n$) and  that in $III$ ($m=n$).
Then, we have to count the number of  times  the path touches the diagonal.

Using  (\ref{ex}),
we can calculate the distribution for $m\leq n$. (See Appendix A.)
\begin{equation}
X(m,n)=
\left\{ 
\begin{array}{ll}
\sum_{k=0}^{m}A_{m,n,k}\frac{p^{n}(1-p)^{m}}{2^{k+1}}& m<n;\\
\sum_{k=0}^{m}A_{m,m,k}\frac{p^{m}(1-p)^{m}}{2^{k}}& m=n.
\end{array} \right \},
\label{dis1}
\end{equation} 
where $A_{m,n,k}$ is given by (\ref{ex}) and 
$k$ is the  number  of  times  the path touches the diagonal.

 The distribution for $m> n$ can be easily  calculated  for the absorption wall $n=m-1$, where $m,n\geq0$.
The distribution for  $m>n$   is given by
\begin{equation}
X(m,n)=\sum_{k=0}^{n}A_{n,n,k}\frac{p^{n}(1-p)^{n}(1-\frac{1}{2}p)}{2^{k+1}} \hspace{2cm} m>n.
\label{dis2}
\end{equation} 

We  investigate  the limit $t\rightarrow \infty$.
Here, we consider $m$ as  a variable;
it is the distribution function of the vote for  $C_1$. 
For large $t$,  we can assume that only the first terms of the summation  of (\ref{dis1}) and (\ref{dis2}) are  non-negligible.
The first term becomes  the difference of the binomial distributions  using (\ref{ex}).
For $m/t<1/2$, the peak of the binomial distribution is $1-p$ and 
for  $m/t>1/2$, it is $1$.
Then, we can obtain the distribution in the scaling limit 
$t=m+n\rightarrow \infty$,
\begin{equation}
\frac{m}{t}\Longrightarrow  Z.
\label{limit1}
\end{equation}
The probability measure of $Z$ is
\begin{equation}
\mu=\alpha \delta_{1-p}+\beta\delta_1,
\label{limit}
\end{equation}
where $\delta_x$ is  Dirac measure.
$Z$ is the ratio of voters who vote to $C_{1}$ from (\ref{limit1}).
The distribution has two peaks, one  at $Z=1$ and the other at  $Z=1-p$.
Now,  we calculate $\alpha$ and $\beta$,
where $\alpha+\beta=1$.
The probability that the path touches the absorption wall $n=m-1$ is given by
\begin{eqnarray}
\beta&=& \sum_{m=0}^{\infty}X(m,m)(1-\frac{p}{2})
=\sum_{m=0}^{\infty}\sum_{k=0}^{m}A_{m,m,k}\frac{p^{m}(1-p)^{m}}{2^{k}}(1-
\frac{p}{2})
\nonumber \\
&=&
(1-\frac{p}{2}) [1+\frac{x}{2}C_0(x)+(\frac{x}{2})^2C_1(x)+(\frac{x}{2})^3 C_2(x)+\cdots ]
\nonumber \\
&=&
(1-\frac{p}{2}) [1+\frac{x}{2}C_0(x)+(\frac{x}{2})^2\{C_0(x)\}^2+(\frac{x}{2})^3\{ C_0(x)\}^{3}+\cdots ]
\nonumber \\
&=&
(1-\frac{p}{2})[\sum_{k=0}^{\infty} \frac{ \{xC_{0} (x)\}^{k}}{2^{k}}]
=
(1-\frac{p}{2})\frac{1}{1-\frac{xC_0(x)}{2}}
\nonumber \\
&=&\frac{4-2p}{3+\sqrt{1-4p(1-p)}}.
\label{beta}
\end{eqnarray}
$C_k(x)$ is the generating function of the  generalized Catalan number (\ref{gcn}),
 and $x=p(1-p)$.
Here, we use the relations (\ref{cnr2}) and (\ref{cnr}).

 \begin{figure}[h]
\includegraphics[width=120mm]{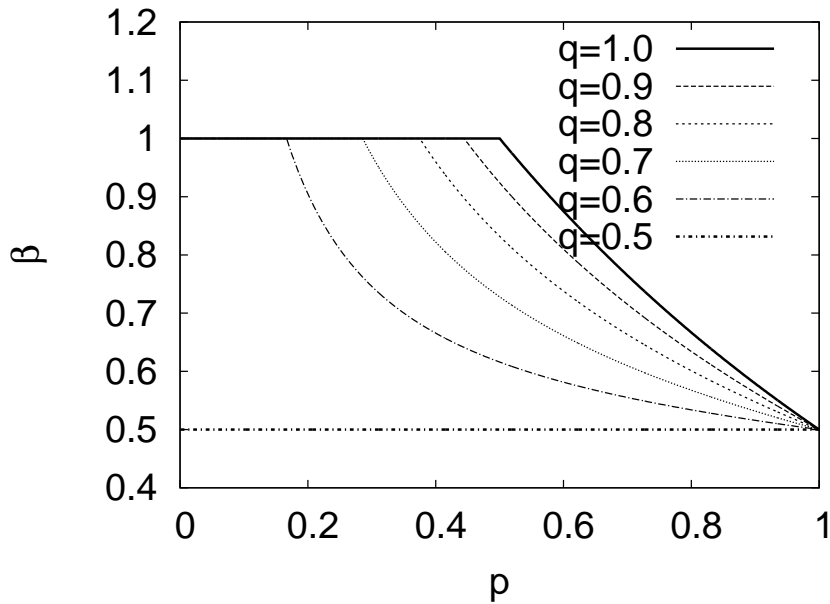}
\caption{$\beta$ or $\bar{s}$, i.e.,  average votes for  candidate  $C_1$ by herders. }
\label{f3}
\end{figure}

In Fig. \ref{f3}, we plot $\beta$   for the case  $q=1$. 
We are interested in the average votes for   $C_1$ by the herders to validate the performance of the  herders.
We define  $s$ as  the average votes for  the correct   candidate $C_1$ by the  herders,
\begin{equation}
s=\frac{ Z-(1-p)}{p}.
\label{s}
\end{equation}
Here, we take expected values of (\ref{s}) about several sequences of voting,
\begin{equation}
\bar{Z}=p\bar{s}+(1-p)=(1-p)\alpha+\beta,
\end{equation}
where $\bar{x}$ means the expected value of $x$.
The second equality can be obtained  from (\ref{limit}).
Using the relation $\alpha+\beta=1$, we can   obtain $\bar{s}=\beta$.

\begin{figure}[h]
\includegraphics[width=120mm]{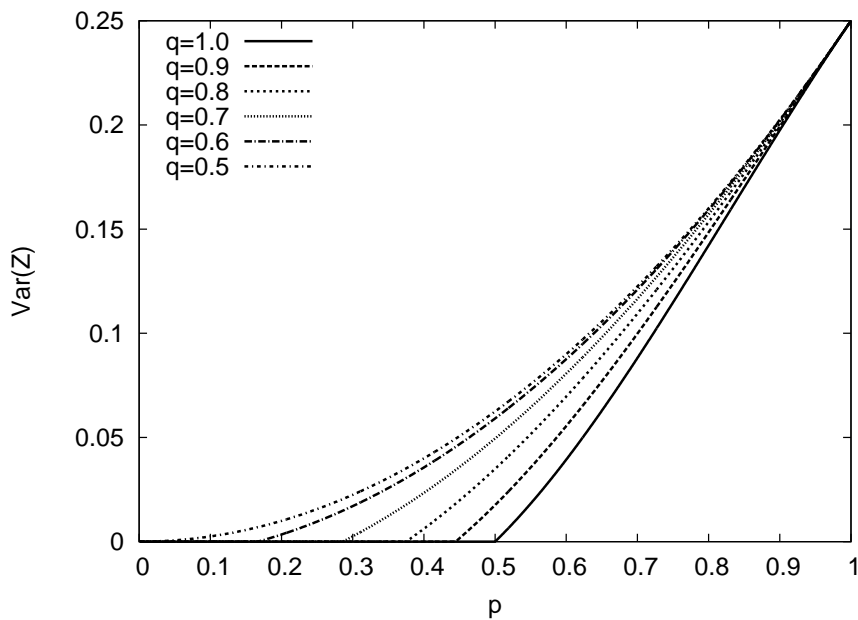}
\caption{$Var(Z)$, the variance of $Z$   is the order parameter. 
$Var(Z)$  is not differentiable at $p_c$. $Z$ is the ratio of voters who vote to $C_{1}$.}
\label{f3.5}
\end{figure}

When $p$ is less than  $0.5$, herding is a highly efficient  strategy.
The distribution of votes peaks when  $Z=1$.
A majority of votes  is  necessary  select the correct candidate $C_1$. 
At $p=p_c=0.5$, there is a phase transition.
When $p$ exceeds $p_c=0.5$, the distribution of votes has  two peaks.
In this case,  a majority  may select the wrong candidate $C_0$.
In  the language of  game theory, this  is a bad equilibrium.
The probability of falling into   bad equilibrium is $1-\beta$.
$Var(Z)$, the variance of $Z$  in the large $t$ limit is the order parameter.
It is observed that $Var(Z)$ is not differentiable at $p=1/2$ (Fig. \ref{f3.5}).
Hence, the phase transition is  of the second order.

When $p\leq p_c$, the distribution  has  one peak, and  it  does not depend on $P_0 (m,m)$, which is the probability of the vote when the number of  votes for  $C_0$ is the same as that for  $C_1$.
We can confirm that $Var(Z)$ is $0$ in Fig. \ref{f3.5}.
On the other hand,  when  $p>p_c$,  
 the  limit distribution  depends on  $P_0 (m,m)$;
$P_{0}(m,m)$ is given by (\ref{p}).
We can confirm that $Var(Z)$ is not $0$ in Fig. \ref{f3.5}.
If herders are analog,  $Var(Z)$ is $0$ in all region of $p$.

Next, we consider the general $q$  case.
In this case, the path goes across the  diagonal several times.
$n=m-1$ is no longer   the absorption wall,  where $m,n\geq0$.
Hence, it is difficult to calculate the exact  solution for general $q$.
However, as in   the discussion of (\ref{limit}),
we can obtain the limit shape of the distribution of the votes for  $C_1$,
\begin{equation}
\frac{m}{t}\Longrightarrow  Z.
\label{limit3}
\end{equation}
The probability measure of $Z$ is
\begin{equation}
\mu=\alpha \delta_{(1-p)q}+\beta\delta_{p+(1-p)q},
\label{limit2}
\end{equation}
where $\alpha+\beta=1$.
When $q=1$, (\ref{limit2}) becomes (\ref{limit}).
We can calculate $\beta$  as
\begin{equation}
\beta=\tilde{R_1}(1-R_2)(1+R_1 R_2+R_1^2R_2^2+\cdots)=\frac{\tilde{R_1}(1-R_2)}{1-R_1R_2}.
\label{general}
\end{equation}

$\tilde{R_1}$ is the probability that  the path  starts from  $(0,0)$,
goes across the diagonal  only once,
 and reaches   the wall $n=m-1$ in  $II$ $(m>n)$.
 $R_1$ is the probability that  the path  starts from  the wall $n=m+1$ in  $I$ $(m<n)$,
goes across the diagonal  only once,
and reaches   the wall $n=m-1$ in  $II$ $(m>n)$. 
$R_2$ is the probability that  the path  starts from  the wall $n=m-1$ in  $II$ $(m>n)$,
goes across the diagonal  only once,
and reaches   the wall $n=m+1$ in  $I$ $(m<n)$. 

For example, the first term of (\ref{general}) is the path that starts from $(0,0)$ and passes  through $I$ $(m<n)$ or directly enters  $II$ $(m>n)$.
The path goes across the diagonal $III$ $(m=n)$  only once;
 the first step is   rightwards.
The second term is  the path that starts from $(0,0)$,  goes  across the diagonal $III$  $(m=n)$ three times, and enters  $II$ $(m>n)$.
We can calculate $\tilde{R_1}$, $R_1$, and $R_2$  similarly to  (\ref{beta}) (See appendix B),
\begin{eqnarray}
\tilde{R_1}&=&\frac{2(1-B)}{2-\gamma_1(1-\sqrt{1-4A(1-A)})},
\nonumber  \\
R_1&=&\frac{(1-B)\gamma_1(1-\sqrt{1-4A(1-A)})}{B\{2-\gamma_1(1-\sqrt{1-4A(1-A)})\}},  \nonumber \\
R_2&=&\frac{B\gamma_2(1-\sqrt{1-4C(1-C)})}{(1-B)\{2-\gamma_2(1-\sqrt{1-4C(1-C)})\}},  
\end{eqnarray}
where $A$, $B$, and $C$ are given by  (\ref{prob}),  $\gamma_1=B/A$, and $\gamma_2=(1-B)/(1-C)$.

In general, we can calculate  the exact  value of  $p_c$:
\begin{equation}
p_c=1-\frac{1}{2q}.
\label{cp}
\end{equation}
As $q$ increases, $p_c$ increases. 
At $p_c$, the model features  a phase transition beyond which  a state where most agents make the correct forecasts coexists with one where most of them are wrong.
Thus, the effectiveness of  herding decreases   as  $q$ decreases.
In the limit  $q=0.5$, the phase transition disappears.
The distribution becomes symmetric in this case.

From the viewpoint of the herders being  noise, if  $p$ is  greater than  $p_c$, the vote ratios deviate   considerably  from the fundamental value $q$.
Thus,  digital herders account for   greater noise  than  analog  herders.
Analog herders affect only the speed of  convergence to the fundamental value.  \cite{Hisakado2}
Independent voters can not oppose  digital herders.


\section{Exact solutions for $r = 1$}

Here,  we discuss the cases $r=1$ besides $p\neq 1$.\footnote{
When $p=1$, all voters are herders, and the distribution becomes the limit shape of beta distribution, as  discussed in \cite{Mori}}
Herders can see  only a vote of the  previous voter.
We define $P_i(t)$ as the probability  that the $(t+1)$th voter votes for  $C_i$,  where $i=0,1$.
Here, $t$ denotes the  time.
\begin{equation}
P_{0}(t)=\left\{ 
\begin{array}{ll}
p+(1-p)(1-q)\equiv F&;Y_{0}(t-1)=1; \nonumber \\
(1-p)(1-q) \equiv G&:Y_{0}(t-1)=0.
\end{array} 
\right \}
\label{pro}
\end{equation}
$Y_{i}(t)=1$ indicates  that at  $t$, the voter votes for  $C_i$.
$Y_{0}(t-1)=1$ indicates  that the previous voter votes for  $C_0$.
On the other hand,  $Y_{i}(t)=0$ indicates     that at  $t$, the voter does not  vote for  $C_i$.
$Y_{0}(t-1)=0$ indicates  that the previous voter votes for  $C_1$.
Thus, $\sum_{l=1}^{t}Y_{i}(l)$ is the total number of  votes for  $C_i$ until  $t$.
Here,  the relation  $P_{1}(t)=1-P_{0}(t)$ holds.
The initial distribution is
\begin{equation}
P_0(0)=
\frac{1}{2}p+(1-p)(1-q).
\label{in}
\end{equation} 

The model was studied as  a one-dimensional correlated random walk \cite{Bohm}\cite{Konno}.
Here, we introduce $X(m,n)$ as the probability distribution.
$m$ is the number of the voters who vote for  $C_{1}$ and
$n$ is the number of the voters who vote  for  $C_{0}$.
The master equation is 
\begin{equation}
X(m,n)=P_{1}(t-1)X(m-1,n)+P_{0}(t-1)X(m,n-1),
\label{master}
\end{equation} 
for $m\geq 0$ and $n \geq 0$, with the initial condition $X(0,0)=1$.

In the limit $t\rightarrow \infty$,
\begin{equation}
X(m,t-m)\Longrightarrow  N(qt,\sqrt{\frac{F(1-G)}{(1-F)G}}t ),
\end{equation}
where $N(\mu,\sigma^2)$ is the normal distribution with mean $\mu$  and   variance $\sigma^2$. (See  Theorem 3.1. in \cite{Bohm})
$F$ and $G$ are given in (\ref{pro}).
Hence, we can obtain the limit shape of the distribution as
\begin{equation}
\frac{m}{t}\Longrightarrow  Z.
\label{r=1}
\end{equation}
The probability measure of $Z$ is
\begin{equation}
\mu=\delta_{q}.
\end{equation}
Then, $r=1$ involves  no phase transition,
 and the majority of the voters do not select the wrong candidate $C_0$.

The  limit distribution is independent of 
 the initial condition $P_0 (0)$ because the distribution $Z$ has only one peak.

\section{Mean field approximation}
We discussed the exact solutions of this model in the cases $r=1$ and $r=\infty$.
A phase transition occurs  in the case $r=\infty$.
On the other hand, there is no phase transition in the case $r=1$.
We must consider the case $1<r<\infty$.
In this section, we analyze  phase transition using mean field approximation.

We define $P_i^{r}(t)$ as the probability of that the $(t+1)$th voter votes for  $C_i$, where $i=0,1$.
The voter can see the previous $r$ voters' votes.
\begin{equation}
P_{0}^{r}(t)=\left\{ 
\begin{array}{ll}
p+(1-p)(1-q)&;\sum_{l=t-r}^{t-1}Y_{0}(l)>r/2; \nonumber \\
p/2+(1-p)(1-q)&;\sum_{l=t-r}^{t-1}Y_{0}(l)=r/2; \nonumber \\
(1-p)(1-q)  &:\sum_{l=t-r}^{t-1}Y_{0}(l)<r/2 \nonumber.
\end{array} 
\right \}
\label {ge}
\end{equation} 
$\sum_{l=t-r}^{t-1}Y_{0}(l)$ gives the total votes for  $C_0$ from  $(t-r)$ to $(t-1)$.
In other words, it is the total number of  votes of the previous $r$ voters  for  $C_0$ at  $t$.
The case $\sum_{l=t-r}^{t-1}Y_{0}(l)=r/2$ appears only  when $r$ is  even.
Here, the relation  $P_{1}^{r}(t)=1-P_{0}^{r}(t)$ holds.
When $r=\infty$, 
(\ref{ge}) reduces  to (\ref{prob}) and
when $r=1$,
(\ref{ge}) reduces to  (\ref{pro}).

We focus on  the probability of the selection of the correct candidate $C_1$.
The distribution of $Z$ is the limit shape of the distribution of votes for  $C_1$.
In general $q$,  we can rewrite the first equality of  (\ref{s}) as 
\begin{equation}
Z=(1-p)q+ps.
\label{1}
\end{equation}
Here, from the viewpoint of the mean field  approximation,  $s$  can be considered  as the sum of the probabilities of
every combination of majorities in the reference  of previous $r$ votes.

Mean field analysis is an  approximation.
We can not obtain   quantitative  conclusions from  this analysis 
owing to   two major reasons.
First,  the analysis does not use  
$P_0(0)$ in (\ref{ge}).
Second, 
this approximation assumes independence of voters, 
which is not true.
These errors are presented   in the next section. 

When  $r$ is odd,
\begin{equation}
s=\sum_{g=\frac{r+1}{2}}^{r}
\left(
\begin{array}{cc}
r\\
g  
\end{array}
\right)
Z^g(1-Z)^{r-g}\equiv\Omega_{r}(Z).
\label{2}
\end{equation}
When  $r$ is even, from the definition of the 
behavior of the herder,
\begin{eqnarray}
s&=&\sum_{g=\frac{r}{2}+1}^{r}
\left(
\begin{array}{cc}
r\\
g  
\end{array}
\right)
Z^g(1-Z)^{r-g}
+
\frac{1}{2}
\left(
\begin{array}{cc}
r\\
r/2  
\end{array}
\right)
Z^g(1-Z)^{r/2}
\nonumber \\
&=&
\sum_{g=\frac{r}{2}}^{r-1}
\left(
\begin{array}{cc}
r-1\\
g  
\end{array}
\right)
Z^g(1-Z)^{r-1-g}=\Omega_{r-1}(Z).
\label{3}
\end{eqnarray}
The even  case $r$  becomes
the odd case $r-1$ from the viewpoint of  mean field analysis.

\begin{figure}[h]
\includegraphics[width=120mm]{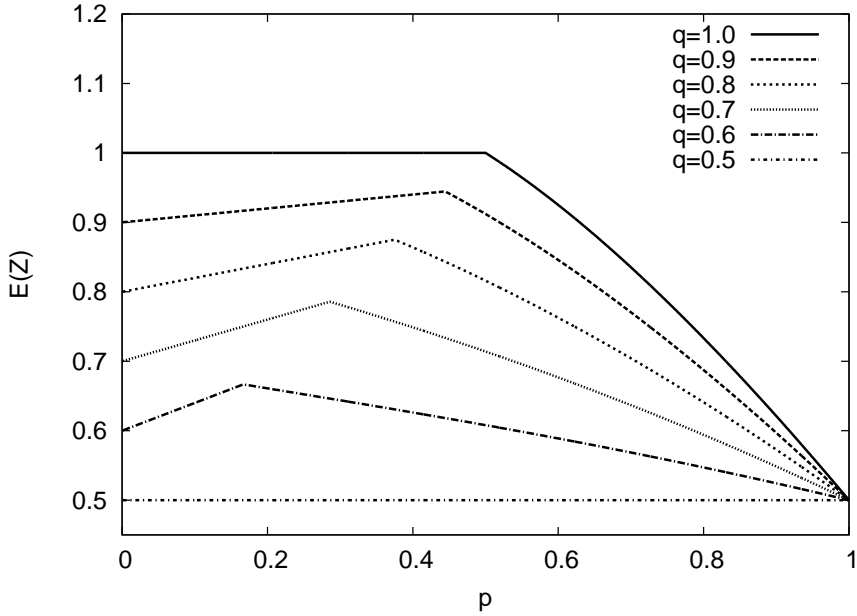}
\caption{Average votes ratio for the correct candidate $C_1$ in the case $r=\infty$. The veritical axis is $E(Z)=\bar{Z}$.  At the critical point $p_c$, $\bar{Z}=E(Z)$ is  maximum.}
\label{f4}
\end{figure}

Fig.  \ref{f4} shows  the exact solutions of $\bar{Z}$, and
Fig. \ref{f3} shows the  exact solutions of $\bar{s}$.
Here $\bar{x}$ is the expected value of $x$.
Both are obtained  from the conclusions of section 3 for  the case $r=\infty$. 
$p$ is the percentage  of herders and $q$ is the
percentage of correct answers of independent voters. 
$\bar{Z}$  increases with  $p$  up to  the critical point (\ref{cp}).
At the critical point, 
 $\bar{Z}$ is   maximum.
Above the critical point, 
 the distribution becomes the  sum of two delta functions, and 
 $\bar{Z}$ decreases  as  $p$ increases.

(\ref{1})  and (\ref{2}) are two self-consistent  equations for $s$ or $Z$.
By substituting  (\ref{2}) in (\ref{1}),
   we obtain
\begin{equation}
s=\Omega_{r}(Z) p+(1-p)q
\label{4}
\end{equation}

\noindent{\bf (1) $r=1$ and $r=2$}

\noindent
In this case, $\Omega_{r}(Z)=q$. We can obtain $s=q$ from (\ref{2}).
Then, we can get $s=Z=q$.
The ratio of herders' votes  for  $C_1$ is constant, as is that of    the independent voters' votes.
There is no transition in these  cases. 
This  is consistent with the  conclusion of section 4.

\noindent{\bf (2) $r=\infty$ }

\noindent(i)$Z>1/2$

\noindent
In this case, $\Omega_{r}(Z)=1$. We can obtain $s=1$ and $Z=p+(1-p)q$.

\noindent(ii)$Z\leq 1/2$

\noindent
In this case, $\Omega_{r}(Z)=0$. We can obtain $s=0$, $Z=(1-p)q$, and the condition $p\geq 1-\frac{1}{2q}$.
Then,  when  $r=\infty$, there is a phase transition at $p_c=1-\frac{1}{2q}$.
When $p\leq p_c$, the herders always vote for the correct  candidate $C_1$.
On the other hand,
 when $p>p_c$, there are two cases.
One is the same as that when  $p\leq p_c$.
In the other  case,  the herders always vote for the wrong  candidate $C_0$.
This phenomenon is known as   an  information cascade.
The conclusion  is consistent with  that of   section 3;
however, by the mean field approximation, 
we can not obtain the exact distributions that  are obtained in  section 3.
 
\noindent{\bf (3) $ 3\leq r<\infty$  }

\begin{figure}[h]
\includegraphics[width=140mm]{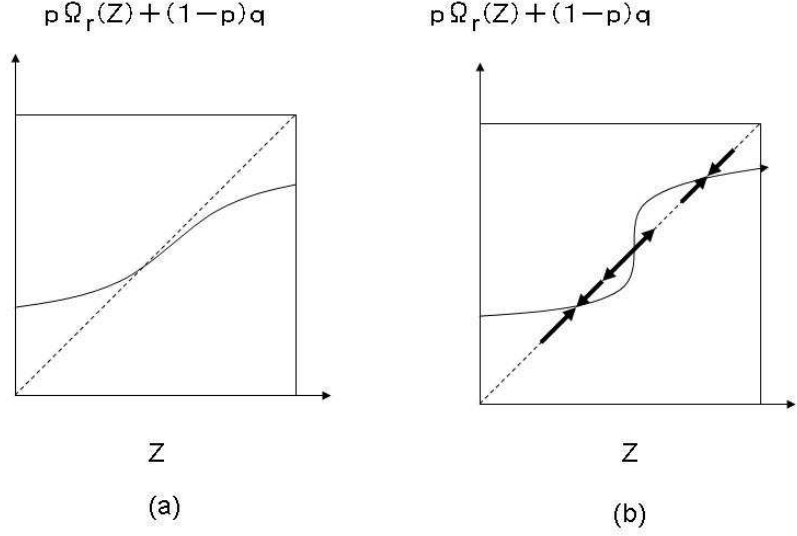}
\caption{Solutions of self-consistent equation (\ref{4}) in the case $ 3\leq r<\infty$  (a) $p\leq p_c$ (b)$p>p_c$. Below the critical point $p_c$ we can obtain one solution (a). On the other hand, above the critical point, we obtain three solutions. Two of them are stable and one is unstable (b). }
\label{f5}
\end{figure}

\noindent
(\ref{4}) admits one solution for  $p\leq p_c(r)$  (see Fig. \ref{f5}(a))
 and  three  solutions for $p>p_c(r)$  (see Fig. \ref{f5}(b)).
When  $p>p_c(r)$, the  upper  and lower solutions 
are  stable solutions;
on the other hand, 
the intermediate solution is  an unstable solution.
Then, the  two  stable solutions attain  good and bad equilibrium,  respectively,
and,  the distribution becomes the sum of the two delta functions,
as  in the case $r=\infty$ (see  section 3).

\section{Numerical Simulations}

In order to confirm the analytical results, we perform numerical simulations.
We adopt two approaches, numerical integration of   the master equation  and Monte Carlo simulation for this model.
The  master equation is given by(\ref{master}) and $P_{i}(t)$ is given by (\ref{ge}).

%
\begin{figure}[h]
\begin{minipage}{.5\textwidth}
\includegraphics[width=\textwidth]{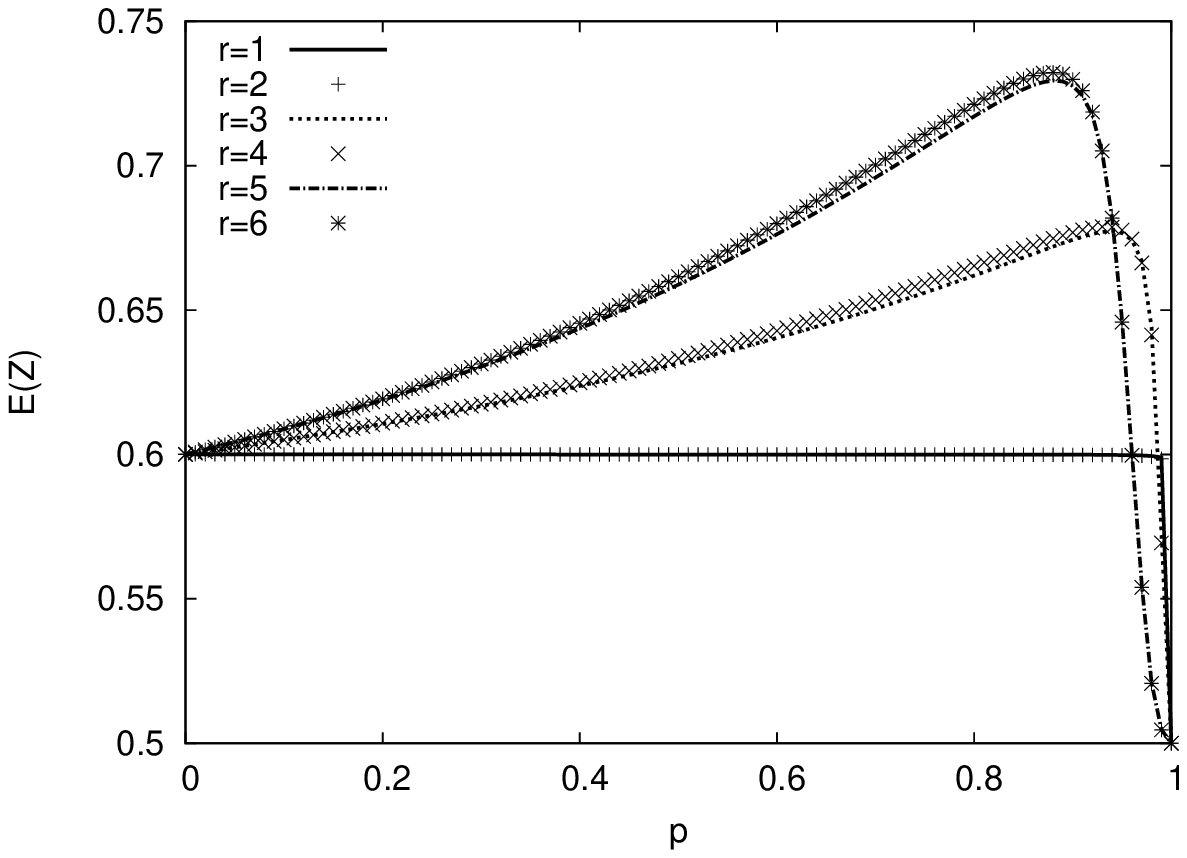}
\end{minipage}
\hspace{0.3cm}
\begin{minipage}{.5\textwidth}
\includegraphics[width=\textwidth]{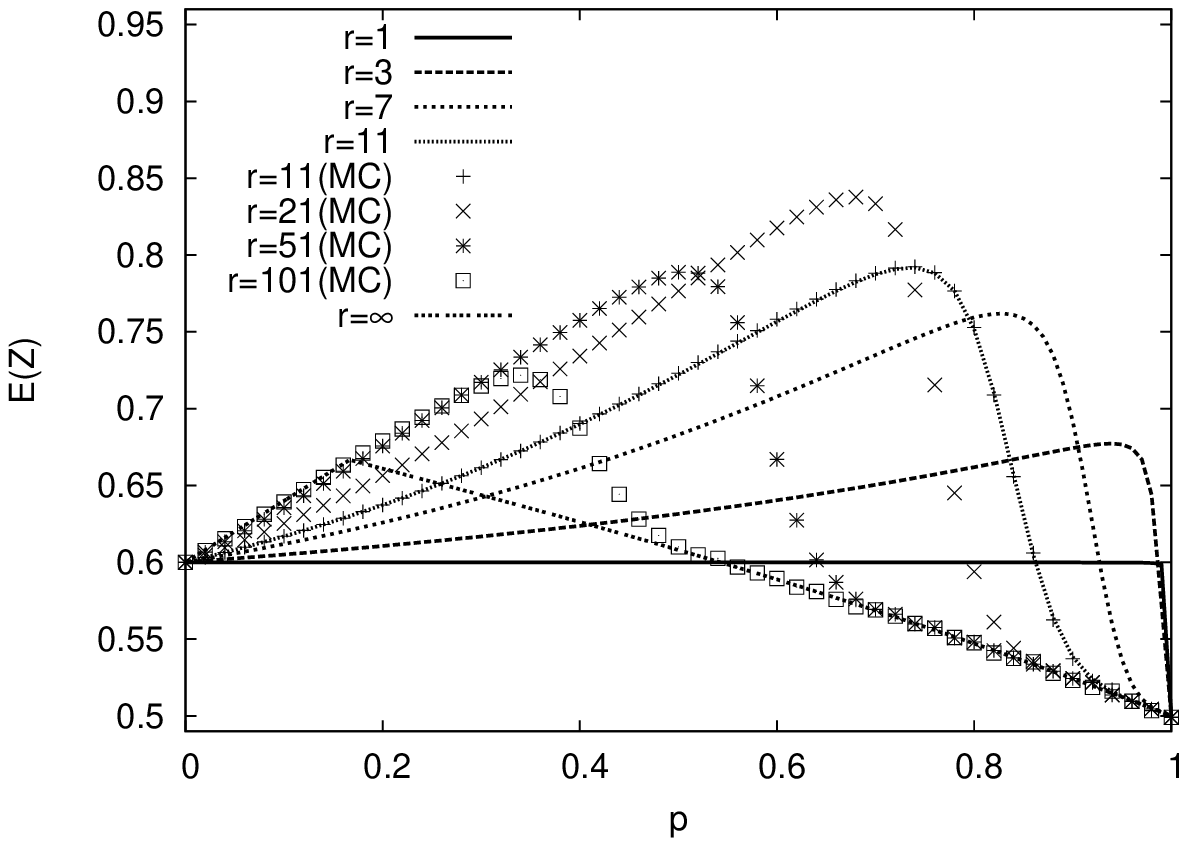}
\end{minipage}
\caption{Average votes ratio for the correct candidate $C_1$ when  $q=0.6$. (a) numerical integration and (b) numerical integration and Monte Carlo simulation.
 The vertical axis is $E(Z)=\bar{Z}$.}
\label{f6}
\end{figure}

Fig. \ref{f6} shows the average votes ratio for the correct candidate $C_1.$ 
Fig. \ref{f6}(a) shows numerical integration at  $t=10000$.
We can see that the  even case $r$   almost coincides with  the odd case $r-1$.
The conclusion based on  the previous section is reasonable (See (\ref{2}) and (\ref{3})). Fig. \ref{f6}(b)  shows  the numerical integration and Monte Carlo simulation. The number  of   simulations is $100000$.
We can check  whether the  Monte Carlo simulation is  consistent with 
the numerical integration.
In Fig. \ref{f6}(b), we can also confirm the exact solution for $r=\infty$.
The case $q=0.6$ in Fig. \ref{f4} corresponds to the case $r=\infty$ in Fig. \ref{f6}(b).
We can observe clearly  the indifferentiable point at $p_c$ in Fig. \ref{f4}.
On the other hand, the point in Fig. \ref{f6}(b) is smoother than the point in Fig. \ref{f4}.
If we increase  the number of Monte Carlo simulations,
the points will appear similar in Fig. \ref{f6}(b) and Fig. \ref{f4}.

Here, we investigate  the maximum $\bar{Z}$,
 i.e., the  maximum probability of  selecting  the correct candidate $C_1$  or maximum  percentage of correct answers.
If the voters can see the previous vote  or two, 
there is no phase transition.
The   percentage of correct answers
  is constant,  as is  the 
 percentage of correct answers of independent voters $q$.
In this case,  the lowest maximum  percentage of correct answers is shown  (Fig. \ref{f6}(b)).
If the voters can see more than  the previous  2 votes,
a phase transition occurs.
Above the critical point $p_c$,
the distribution has  two peaks.
As $r$ increases, the critical point $p_c$ decreases.

When the voters can see  all the previous   votes,
we believe that  the   maximum  percentage of correct answers
is  the highest.
For example, when  we select the herding strategy, we  collect as much information as we can. 
However, this  is not true.
When  the voters can see the   previous 21 votes,
the   maximum  percentage of correct answers
is the highest when $q=0.6$.
Too much information induces mistakes among  the herders.
It can be observed  in   collective behaviors  of animal groups  such as fish schools and bird flocks \cite{frank}.
There may be limits to the information available to grouping individuals.
The average distance maintained between neighbors within pelagic fish schools is usually between three-tenths of body length and one body length.
Individuals  can change their position relative to others only  on the basis of   local information.
They do not need information of the  entire group.

The maximum probability of the selecting  the  correct  answer is at the critical point $p_c$  when  $r=\infty$. This can be seen  in Fig. \ref{f4}, which is the exact solution for  the case  $r=\infty$.
On the other hand, for  $3\leq r<\infty$,  the maximum probability of the selecting  the  correct answer is above the critical point $p_c$. In this phase, the distribution of  votes has two peaks. Thus, the possibility of the majority of  voters  selecting  the wrong answer increases; however, the average probability of selecting the  correct answer increases.
We discussed the unstable solution  in the  2 peaks phase in the previous section.

As discussed in the previous section, we  find that  the conclusions  of numerical integration and Monte Carlo simulation  are  inconsistent with the conclusion of  mean  field approximation analysis in some respects.
For example, when  $r=3$ and $q=0.6$, the critical point  is at  $p_c=0.78$ from  mean field analysis. At  this point,  we get  $\bar{Z}=0.89$ by  using this method. On the other hand,  from the conclusion of the numerical simulations, we get  $p_c=0.74$ and $\bar{Z}=0.65$ (Fig. \ref{f6}(b)). 
The rough estimate of  the  critical point $p_c$ can be calculated  by  mean field approximation; however, it   is difficult to estimate $\bar{Z}$.
Mean field approximation is excessively optimistic because it  does not use the information at $t=1$ and $\sum_{l=t-r}^{t-1}Y_{0}(l)=r/2$. When $r=1$ and $r=2$,  the distribution $Z$ has one peak; hence, it is independent of  these conditions. Thus, the conclusions   from  mean fields approximation  are consistent with those from  numerical simulations. On the other hand,  when $3\leq r<\infty$, the distribution $Z$ depends on  these conditions. 

\section{Social experiments}

We conducted simple social experiments for our model.
We framed  200 questions, each  with two choices--
knowledge and  no knowledge.
31 participants answered  these questions sequentially. 
First, they answered the questions without any information about the others'  answers,
 i.e., their  answers were   based on  their own knowledge.
Those  who   knew the answers    selected the correct answers.
Those who    did not  know the answers   selected the correct answers  with  a probability of $0.5$. 
Next,  the  participants were allowed to see   the  previous participants' answers.
Those who  did  not  know the answers  referred  to this  information.
We are interested in whether they referred to the  information as digital herders or  analog herders.

\begin{figure}[h]
\begin{minipage}{.5\textwidth}

\includegraphics[width=\textwidth]{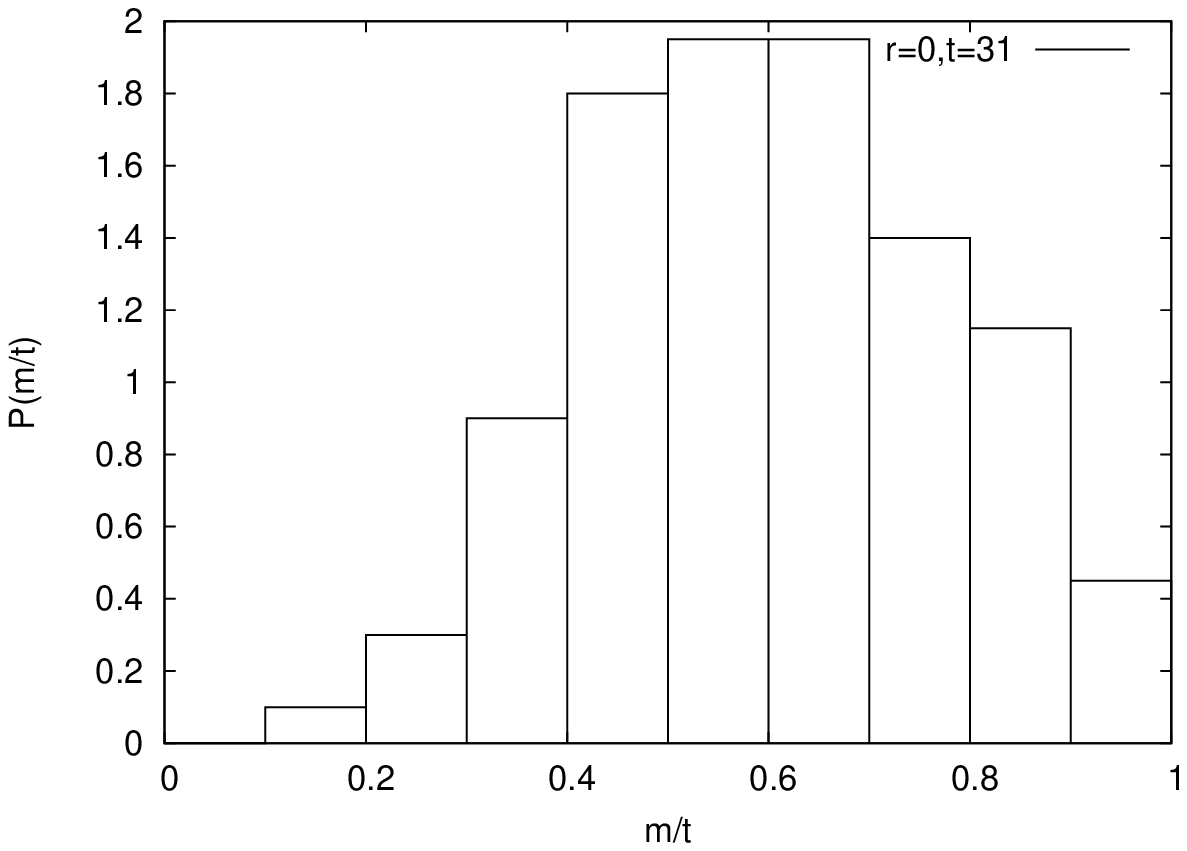}
\end{minipage}
\hspace{0.3cm}
\begin{minipage}{.5\textwidth}
\includegraphics[width=\textwidth]{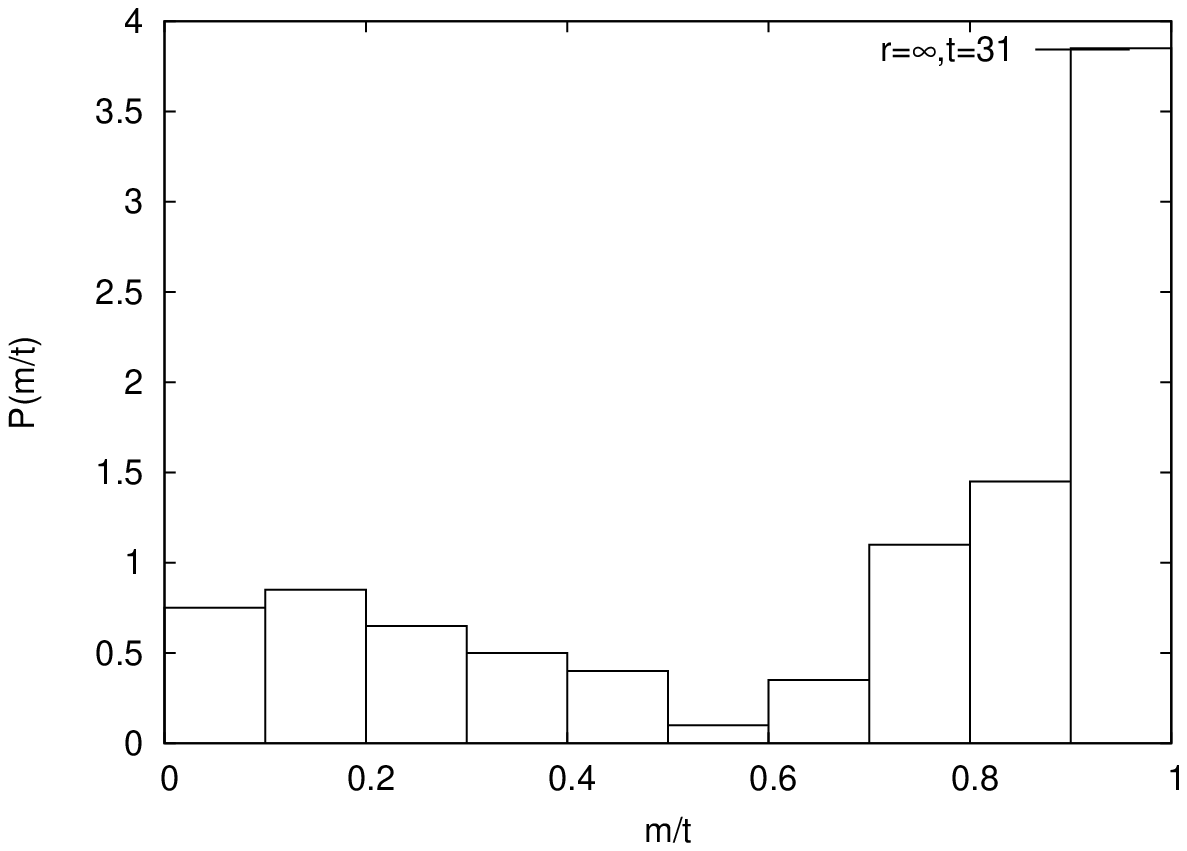}
\end{minipage}
\caption{Distribution of correct  answers. (a) $r=0$ and   (b) $r=\infty$.
 We can observe  one peak at $Z=m/t=0.6$ in (a). We can observe  two peaks, one  at $Z=m/t=0.2$ and  the other  at $Z=m/t=1$ in (b).  The peak at $Z=0.2$ is attributed to the wrong answers caused by  the information cascade.}
\label{f7}
\end{figure}

Fig. \ref{f7} shows the results of the social  experiments,
 i.e., the distribution of  the correct answers:
(a) $r=0$ and (b)  $r=\infty$.
The average  correct answer ratio is $0.6$ in the case of (a).
Hence, the independent voters who knew the answers account for  $0.2$ of all voters,  and 
the herders  who did know the answers account for    $0.8$ of all the voters.
If the herders are digital, we can apply the model described  in section 3, with $p=0.8$, $q=1$, and $r=\infty$.
From (\ref{limit}),  the distribution of percentage of   correct answers  has    peaks at $Z=m/t=0.2$ and $Z=m/t=1$.
Although the number of  votes is  small and does not converge, the prediction can be recognized in Fig. \ref{f7} (b).
The peak at $Z=0.2$ is attributed  to wrong answers  caused by  rational choices.
This phenomenon is known as  is an information cascade;
it is  caused  by digital herders and  not by  analog herders.
If the herders are analog herders, there is one peak at $Z=1$ and there is no information cascade in the case of Fig. \ref{f7}(a).  
We believe that  in this case,   almost all herders behave as digital herders.  

\section{Concluding Remarks}

\begin{figure}[h]
\includegraphics[width=120mm]{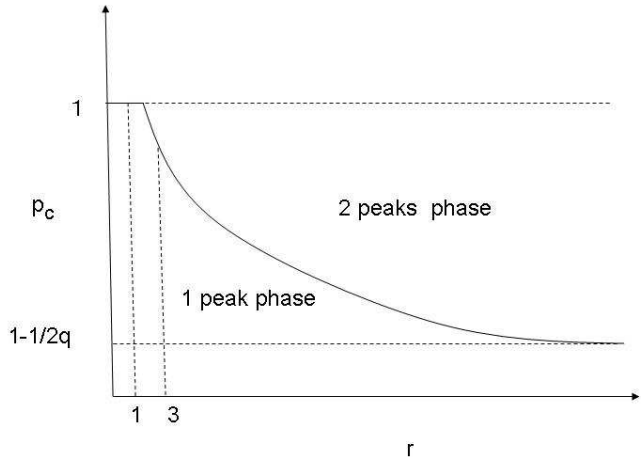}
\caption{Phase diagram in  space $p_c$ and $r$. $r$ is  an integer. When  $r=1$ and $r=2$,   there is no phase transition. When $r\geq 3$,  there is a  phase transition. There are two phases. One is a one peak phase, the other is two peaks phase. Two peaks phase represents  the information cascade. As $r$ increases, the critical point  decreases. 
}
\label{f8}
\end{figure}

We investigated a voting model  that is similar to 
a Keynesian beauty contest.
We calculated the exact solutions for the   special cases $r=1$ and 
$r=\infty$, 
and
we analyzed the general case using mean field approximation and numerical simulations.
When  $r=1$ and $r=2$,   there is no phase transition.
The percentage of  correct answers is the same as that
of  independent voters.
In this case,  herders can not increase
the percentage of  correct answers.
When $r\geq 3$,  there is  phase transition.
As the fraction of herders increases, the model features  a phase transition beyond which  a state where most voters make the  correct votes coexists with one where most of them are wrong.
As $r$ increases, the critical point  decreases. 
The phase diagram is shown  in Fig. \ref{f8}.
When $r=\infty$, 
we can obtain the exact  solutions.
When $r\leq 3<\infty$,
we can not obtain the critical point $p_c$ and the distribution precisely.
In this   case,  mean  field approximation analysis is  inadequate.
It is  a problem that must be addressed in the future. 

The high critical point 
induces a  low risk of  phase transition.
Ants use chemical signals called pheromones,
and they  behave as herders.
The pheromones evaporate quickly.
As an analogy, in our model,   $r$ is small.
Thus, pheromones  may amplify the limited intelligence of  individual ants into something more powerful  to avoid phase transition.


We are  also interested in the behavior of human beings.
We conducted simple experiments for our model when $r=0 $ and $r=\infty$.
Although the total number of votes is small and does not converge, in these experiments  an information cascade is observed.
This phenomenon was caused  by digital herders and  not by  analog herders.
If the herders are analog,  the difference of the phase is only the velocity of the convergence.
Analog herders do not lead to erroneous decisions in $t=\infty$.
On the other hand, 
 if the  herders are digital   
the distribution of the votes has two peaks.
One represents good equilibrium  and the other represents  bad equilibrium.
In the case of bad equilibrium, herders make  erroneous decisions at $t=\infty$.
We can conclude that the information cascade  is caused by  the phase transition of  digital herders.
 Detailed analysis of the  experiments is a problem that must be addressed in the future.

Finally,  we comment on  the relations between our model and the model 
 introduced by Curty and Marsili \cite{Curty}.
The mean field equations of their model are  same as the  (\ref{1}),  (\ref{2}), and (\ref{3}).
The difference is as follows.
(1) The number of the agents, $N$, of their model is finite.
(2) The interactive process is repeated until  it converges.
We are interested  in  (2), i.e.,
the interaction among the voters.
In the future, we plan to investigate 
 the effects of the interaction on the distributions of votes.

\appendix\section*{Acknowledgment}
This work was supported by
Grant-in-Aid for Challenging Exploratory Research 21654054 (SM).

\def\thesection{Appendix \Alph{section}}

\section{Catalan Number}

Here, we consider the number of monotonic paths along the edges of a grid with square cells, which do not pass lower the diagonal.
Let $m$ and $n$  be the horizontal axis and  the vertical axis, respectively.
The coordinates of the lower left corner are  $(0,0)$.
A monotonic path is one which starts in the lower left corner, finishes in the upper triangle $(m,n)$,  where $0\leq m\leq n$, and consists entirely of edges pointing rightwards or upwards.

The number of  paths  from $(0,0)$ to $(m,n)$ can be calculated as
\begin{equation}
C_{m,n}=\frac{(n-m+1)(n+m)!}{m!(n+1)!}
=
\left(
\begin{array}{cc}
n+m\\
n  
\end{array}
\right)
-
\left(
\begin{array}{cc}
n+m\\
n+1  
\end{array}
\right)
.
\label{gc}
\end{equation}
These numbers  are known as generalized  Catalan number.

If the finish point is $(m,m)$, the number of paths becomes 
the Catalan  number
\begin{equation}
C_{m,m}=c_m
=
\frac{2m!}{m!(m+1)!}
=
\left(
\begin{array}{cc}
2m\\
m  
\end{array}
\right)
-
\left(
\begin{array}{cc}
2m\\
m+1  
\end{array}
\right)
.
\end{equation}

Next,  we compute the distribution of  the number of  the paths that  start in the lower left corner, finish in the upper triangle $(m,n)$,  and  touch the  diagonal $k$ times \cite{fg}.
Let $A_{m,n,k}$ denote the number of paths that   touch the diagonal $k$ times.
We get a simple recursion relation about $A_{m,n,k}$,
\begin{equation}
A_{m,n,k}=\sum_{j=0}^{m-1}c_{j}A_{m-j-1,n-j-1,k-1},
\label{rr}
\end{equation} 
for $k\geq 0$, $n,m\geq 0$, and $m\geq k$, with  the initial condition  $A_{0,0,0}=1$.
This  defines the numbers $A_{m,n,k}$ uniquely, and it is easy to prove that
\begin{eqnarray}
A_{m,n,k}&=&\frac{(n-m+k)(n+m-k-1)!}{n!(m-k)!} \nonumber \\
&=&
\left(
\begin{array}{cc}
n+m-k-1\\
n-1  
\end{array}
\right)
-
\left(
\begin{array}{cc}
n+m-k-1\\
n 
\end{array}
\right)
.
\label{ex}
\end{eqnarray}  
From (\ref{gc}) and (\ref{ex}), we can obtain the relation:
\begin{equation}
A_{m,m,k}=C_{m-k,m-1}.
\label{re}
\end{equation}

The  well known generating function $C_0(x)$  of Catalan numbers is given  by 
\begin{eqnarray}
C_0(x)&=& \sum_{n=0}^{\infty}C_{m,m}x^{n}
\nonumber \\
&=& 1+x+2x^2+5x^3+14x^4+42x^5+132x^6+\cdots,
\end{eqnarray}
subject to the algebraic relation
\begin{equation}
xC_0(x)^2=C_0(x)-1,
\end{equation}
and we can obtain
\begin{equation}
C_0(x)=\frac{1-\sqrt{1-4x}}{2x}.
\label{cnr2}
\end{equation}
Here, we  obtain the generating function $A_{m,k}(x)$ of $A_{m,m,k}$ .
\begin{eqnarray}
A_{m,k}(x)&=& \sum_{m-k=0}^{\infty}A_{m,m,k}x^{m-k}
=\sum_{m-k=0}^{\infty}C_{m-k,m-1}x^{m-k}
\nonumber \\
&=&\sum_{l=0}^{\infty}C_{l,l+k-1}x^{l}
=C_{k-1}(x).
\label{gcn}
\end{eqnarray}
We use (\ref{re}) for the second equality.
$C_{j}(x)=\sum_{l=1}^{\infty}C_{l,l+j}x^{l}$ is the generating function of 
 the generalized  Catalan number (\ref{gc}).
The generating function of 
 the generalized  Catalan number is given by
\begin{eqnarray}
C_1(x)&=& \sum_{n=0}^{\infty}C_{m,m+1}x^{n}
\nonumber \\
&=& 1+2x+5x^2+14x^3+42x^4+132x^5+429x^6+\cdots,
\nonumber \\
C_2(x)&=& \sum_{n=0}^{\infty}C_{m,m+2}x^{n}
\nonumber \\
&=& 1+3x+9x^2+28x^3+90x^4+297x^5+1001x^6+\cdots,
\nonumber \\
C_3(x)&=& \sum_{n=0}^{\infty}C_{m,m+3}x^{n}
\nonumber \\
&=& 1+4x+14x^2+48x^3+165x^4+572x^5\cdots.
\nonumber 
\end{eqnarray}
From (\ref{rr}), 
we can obtain
\begin{equation}
C_{j}(x)=C_{j-1}(x)C_{0}(x).
\end{equation}
Thus, the simple relation between  the generating functions is given by
\begin{equation}
C_{j}(x)=\{ C_{0}(x) \}^{j+1}.
\label{cnr}
\end{equation}

\section{Derivation of $\tilde{R_1}$, $R_{1}$, and $R_{2}$}

\begin{figure}[h]
\includegraphics[width=120mm]{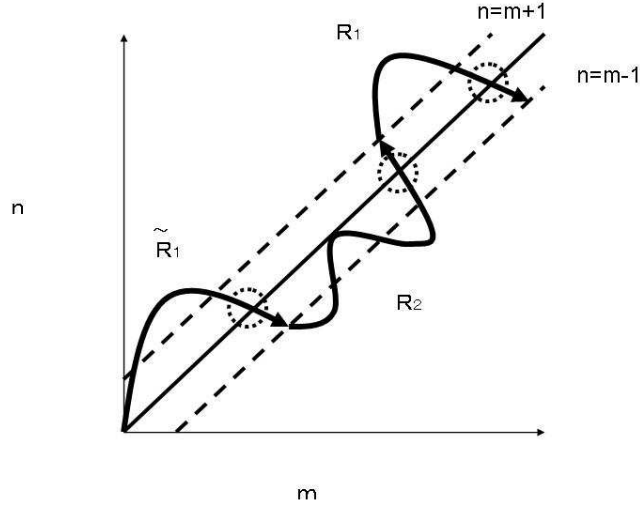}
\caption{$\tilde{R_1}$, $R_1$, and $R_2$. $\tilde{R_1}$ is the probability that  the path  starts from  $(0,0)$,
goes across the diagonal  only  once,
 and reaches   the wall $n=m-1$ in  $II$ $(m>n)$. 
$R_1$ is the probability that  the path  starts from  the wall $n=m+1$ in  $I$ $(m<n)$,
goes across the diagonal  only once,
and reaches   the wall $n=m-1$ in  $II$ $(m>n)$. 
$R_2$ is the probability that  the path  starts from  the wall $n=m-1$ in  $II$ $(m>n)$,
goes across the diagonal  only once,
and reaches   the wall $n=m+1$ in  $I$ $(m<n)$.  
}
\label{f9}
\end{figure}

$\tilde{R_1}$ is the probability that  the path  starts from  $(0,0)$,
goes across the diagonal only once,
 and reaches   the wall $n=m-1$ in  $II$ $(m>n)$  (Fig. \ref{f9}).
\begin{eqnarray}
\tilde{R_1}&=& 
(1-B) [1+y\gamma_1 C_0(y)+(y\gamma_1)^2C_1(y)+(y\gamma)^3 C_2(y)+\cdots ]
\nonumber \\
&=&
(1-B) [1+y\gamma_1 C_0(y)+(y\gamma_1)^2\{C_0(y)\}^2+(y\gamma_1)^3\{ C_0(y)\}^{3}+\cdots ]
\nonumber \\
&=&
(1-B)[\sum_{k=0}^{\infty} \{\gamma_1 yC_{0} (y)\}^{k}]
=
\frac{1-B}{1-\gamma_1 AC_0(A)}
\nonumber \\
&=&\frac{2(1-B)}{2-\gamma_1(1-\sqrt{1-4A(1-A)})},
\label{gamma}
\end{eqnarray}
where $A$ and  $B$ are given by  (\ref{prob}),  $\gamma_1=B/A$, and $y=A(1-A)$.
$C_k(y)$ is the generation function of the  generalized Catalan number (\ref{gcn}).
Here, we use the relations (\ref{cnr}) and (\ref{cnr2}).
When $q=1$,  (\ref{gamma}) reduces to (\ref{beta}).

$R_1$ is the probability that  the path  starts from  the wall $n=m+1$ in  $I$ $(m<n)$,
goes across the diagonal  only once,
and reaches   the wall $n=m-1$ in  $II$ $(m>n)$. 
\begin{eqnarray}
R_1&=& 
\frac{1-B}{B} [y\gamma_1 C_0(y)+(y\gamma_1)^2C_1(y)+(y\gamma_1)^3 C_2(y)+\cdots ]
\nonumber \\
&=&
\frac{1-B}{B}[y\gamma_1 C_0(y)+(y\gamma_1)^2\{C_0(y)\}^2+(y\gamma_1)^3\{ C_0(y)\}^{3}+\cdots ]
\nonumber \\
&=&
\frac{1-B}{B}[\frac{\tilde{R_1}}{1-B}-1]
=
\frac{(1-B)\gamma_1(1-\sqrt{1-4A(1-A)})}{B\{2-\gamma_1(1-\sqrt{1-4A(1-A)})\}}.
\label{gamma1}
\end{eqnarray}

$R_2$ is the probability that  the path  starts from  the wall $n=m-1$ in  $II$ $(m>n)$,
goes across the diagonal  only once,
and reaches   the wall $n=m+1$ in  $I$ $(m<n)$. 
\begin{eqnarray}
R_2&=& 
\frac{B}{1-B} [z\gamma_2 C_0(z)+(z\gamma_2)^2C_1(z)+(z\gamma_2)^3 C_2(z)+\cdots ]
\nonumber \\
&=&
\frac{B}{1-B}[z\gamma_2 C_0(z)+(z\gamma_2)^2\{C_0(z)\}^2+(z\gamma_2)^3\{ C_0(z)\}^{3}+\cdots ]
\nonumber \\
&=&
\frac{B\gamma_2(1-\sqrt{1-4C(1-C)})}{(1-B)\{2-\gamma_2(1-\sqrt{1-4C(1-C)})\}}.
\label{gamma2},
\end{eqnarray}
where $C$ is  given by  (\ref{prob}),  $\gamma_2=(1-B)/(1-C)$, and $z=C(1-C)$.

\end{document}